\documentclass[%
 reprint,
 amsmath,amssymb,
prb,
]{revtex4-2}
\pdfoutput=1
\usepackage{graphicx}
\usepackage{dcolumn}
\usepackage{bm}
\usepackage{float}
\usepackage{pdfpages}
\usepackage{etoolbox} 
\usepackage{pgffor}

\makeatletter
\AtBeginDocument{\let\LS@rot\@undefined}
\begin{document}

\preprint{APS/123-QED}

\title{Polysaccharide conformations measured by solution state x-ray scattering}
\author{Bradley W. Mansel}
\affiliation{Department of Chemical Engineering, National Tsing Hua University, Hsinchu 30013, Taiwan}
\author{Timothy M. Ryan}
\affiliation{Australian Synchrotron, Clayton, VIC 3168, Australia}
\author{Hsin-Lung Chen}
\affiliation{Department of Chemical Engineering, National Tsing Hua University, Hsinchu 30013, Taiwan}
\author{Leif Lundin}
\affiliation{RISE, Agrifood and Bioscience, Gothenburg, Sweden}
\author{Martin A.K. Williams}
\affiliation{School of Fundamental Sciences, Massey University, Palmerston North 4442, New Zealand}
\affiliation{The MacDiarmid Institute for Advanced Materials and Nanotechnology, Wellington 6140, New Zealand}





\date{\today}


\begin{abstract}
Polysaccharides are semi-flexible polymers composed of sugar residues with a myriad of important functions including structural support, energy storage and immunogenicity.  The local conformation of such chains is a crucial factor governing their interactions, where the relative orientation of adjacent sugar rings determines the propensity for hydrogen bonding and specific ion-mediated interactions with neighbouring chains. Traditionally this conformation has only been directly accessible in the solid-state, using crystallographic techniques such as fibre diffraction.  Herein it is demonstrated that improvements in the quality of synchrotron-based x-ray scattering data means that conformation-dependent features, the positions of which are related to the linear repeating distance between single saccharide monomers, can now be measured in solution. This technique is expected to be universally applicable for polysaccharides that consist of comparatively stiff glycosidic linkages, and to have extensive relevance for a number of biological macromolecules, including glycosylated proteins. 

\end{abstract}

\maketitle

Polyuronates are a class of charged polysaccharides that play crucial roles in the physiology of both land and marine plants (pectic polymers in the cell walls of land plants contain large amounts of polygalacturonic acid, while brown seaweeds contain alginic polymers comprising of polyguluronic and polymannuronic acids). Understanding the structure of these important polysaccharides is not only crucial for appreciating their biological functionality and how this biomass is exploited, but also understanding key steps in the evolution of life on our planet \cite{hobbs2016kdgf}.

\begin{figure}
\includegraphics[width=1\columnwidth]{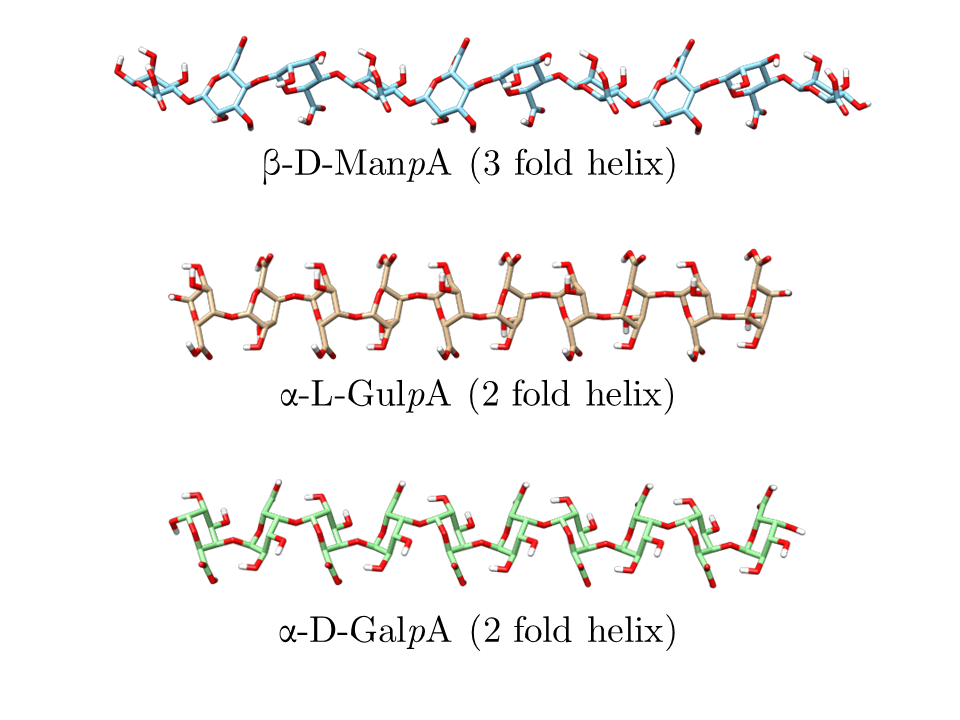}
  \caption{Real-space atomistic models showing the free energy minimizing conformations for the three different conformers used in calculating the scattering presented in figures \ref{fgr:alginate} and \ref{fgr:galac}. Dihedral angles were obtained from ref. \cite{braccini1999conformational} which utilized molecular mechanics calculations. }
  \label{fgr:conformers}
\end{figure}

 A key aspect of the molecular structure of these polymers is the local helical conformation of the chain, which is dictated by the flexibility of the glycosidic linkage between individual sugar residues, and its response to environmental conditions such as pH, ionic strength, specific ion effects and moisture content. These conformations (e.g 2(1) or 3(1) helices) relate to rotations of consecutive sugar residues along the chain and can be thought of as being characterized by the number of residues after which the local structure appears identical when viewed from the end of the molecule, see figure \ref{fgr:conformers}. The orientation of one sugar ring relative to the next along the backbone of multi-residue saccharide chains is determined by the energetics of the glyosidic linkage. Free energy calculations as a function of two angles that uniquely define the spatial relation of one ring to the next typically show several distinct local minima that yield the atomic structures of the maximally populated states of the chain conformation, for example 2(1) or 3(1) helices. The preferred local helical conformation is a function of environmental conditions, such as pH, that modulate the charged nature of the rings and is, amongst other things, key to the propensity of chains to interact with one another to form higher order assemblies. Indeed, the conformation is crucial in determining how the chains interact with each other (including when mediated by specific ions), how chains might respond to mechanical stresses, and how in solution they may interact with other biological macromolecules, such as the enzymes that process them. 
 
 Although many solution-based techniques, such as optical rotation measurements, have been applied to observe changes between local helical conformations, the direct measurement of the atomistic structures of the conformational states has, to date, relied on techniques such as fiber diffraction. Direct experimental measurement of these helical conformations has been carried out solely in the solid state, in fibres or gels \cite{palmer1945x, wuhrmann1945optik, walkinshaw1981conformations, atkins1973structural,atkins1973structural1, mackie1971conformations}, or in crystals formed from oligomeric components \cite{rigby2000observations}, using x-ray diffraction techniques. Indeed there is a rich history of the determination these states and their sensitivity to different preparation conditions \cite{morris1982conformations, li2007reexamining}. 

 As the force fields used to model these molecules in molecular dynamics simulations \cite{peric2008conformation,plazinski2012molecular, panczyk2018extension} are refined, the correct prediction of these helical conformational states in solution (or when bound to chelating ions) is a crucial requirement \cite{wolnik2013probing}. However, at present predictions can only be compared with experimental results that have been measured in the condensed state. While for simulations of multiple chains binding ions between them, commonly carried out in an attempt to clarify aggregation mechanisms that underlie phenomena such as gelation, this seems reasonable \cite{mackie1983aspects}, direct comparison with solution state data has been lacking. 

Here for the first time it is shown that careful x-ray scattering experiments using synchrotron radiation can directly reveal the helical conformation of polysaccharide chains in the solution state. While many SAXS studies have been carried out on pectin and alginate solutions and gels, the focus has been on attempting to understand the formation of the so-called junction zones \cite{sikorski2007evidence, li2007reexamining} (stress bearing connections that build the polymers into a gel network) and their chain multiplicity \cite{stokke2000small, yuguchi2000small, yuguchi2016local, schuster2011using} or on attempts to fit the data to the predictions of statistical physics models of chain configuration of varying degrees of complexity \cite{ventura2013insights, josef2012conformation}. In this work by performing experiments on different oligosaccharides that comprise of varying types of sugar residues, have different inter-ring linkages (di-axial or di-equatorial), and are therefore predicted to exhibit different helical conformations in the solution state, we conclusively show that it is now possible at modern synchrotron facilities to obtain sufficient signal to noise ratios, even at traditionally-ignored high q values, to directly reveal the helical conformation of polysaccharide chains in the solution state. Such high q values would traditionally be thought of as WAXS, and would not typically be recorded concurrently in routine biological solution SAXS experiments. Additional comparison of the results of calculation and experiment for several different degrees of polymerisation lends further weight to our contention regarding the origin and usefulness of the observed high-q peaks.

Oligomeric forms of the two components of alginate, $\beta$-D-Mannopyranuronic acid ($\beta$-D-Man\textit{p}A) and $\alpha$-L-Gulopyranuronic acid ($\alpha$-L-Gul\textit{p}A), with degrees of polymerisation (dp) of 5, 10 and 20 were purchased from Elicityl (http://www.elicityl-oligotech.com). Samples were dissolved, as received, in ultra pure water (18.2 M$\Omega$) to give a final concentration of 50 g/L. Oligomeric pectic acid, $\alpha$-D-Galactopyranose ($\alpha$-D-Gal\textit{p}A) with dps of 6 and 10 were homemade (see ref. \cite{mansel2019resolving} or the Supplemental Material \cite{sup_mat}) and were dissolved in ultra pure water to give final concentrations of 12.5 and 15 g/L, respectively. A sample of $\alpha$-D-Gal\textit{p}A containing a mixture of dps between 25-50 purchased from Elicityl (referred to as OGA) was prepared identically to give a concentration of 31 g/L. The concentrations were selected in order to maximise the signal while optimising the amount of experiments that could be carried out given the amount of sample available. Between them this sample set spans components of the key anionic polysaccharides alginate and pectin.

The small and wide angle X-ray scattering beamline at the Australian Synchrotron was utilized to produce monochromatic X-rays of 15 keV and irradiate samples (multiple 1 second durations) contained in a 1.5 mm path-length solution flow cell running at a rate of 5 $\mu$L/sec \cite{kirby2013low}. Scattered radiation was collected on a Pilatus 1M (Dectris, Switzerland) detector located 0.7379 m from the sample. Background and sample measurements were performed under identical conditions with integration and reduction performed using the standard beamline protocols and the measured intensity put onto an absolute scale using water as a standard \cite{kirby2013low}. The magnitude of the scattering vector is defined as $q=|\mathbf{q}|= 4\pi \lambda^{-1} \sin( \theta /2)$ where $\lambda$ is the wavelength of the incident beam and $\theta$ the angle between the incident beam and the scattered radiation.

During a solution state X-ray scattering experiment the measured scattering, after background subtraction, is a combination of amplitudes from the molecule of interest in-vacuo $A_a(q)$, a boundary layer of water $A_b(q)$ which will have a different density to bulk water, and an excluded volume of water $A_c(q)$ \cite{svergun1995crysol}. The three components are related by
\begin{equation}
I(q) = \langle \mid A_a(q)-\rho_0A_c(q)+\delta \rho A_b(q) \mid^2 \rangle_\Omega    
\end{equation}
where $\rho_0$ is the density of the bulk solvent, $\delta\rho$ is the difference in density between the bulk and boundary layer solvent and $\Omega$ indicates an average over solid angle in reciprocal space \cite{svergun1995crysol}. The small angle scattering for $q<0.8$ \AA$^{-1}$ from a molecule in solution can be routinely calculated with a number of different software packages. While scattering at $q \geq 0.8$ \AA$^{-1}$ (wide angle regime) is significantly more difficult to model due to the length-scales approaching the first peak in the water structure factor, which makes accurately accounting for $\delta \rho A_b(q)$ difficult. Here we used two methods to ensure that the calculated features were independent of the specific software utilized and the particular implementation used in accounting for boundary water and excluded solvent volume. Firstly, calculations were performed using a recent version of the popular software CRYSOL, CRYSOL 3.0 \cite{franke2017atsas}, which incorporates the three necessary amplitudes and, importantly measurements at $q>0.8$ \AA$^{-1}$, of dummy water molecules for use in $A_b(q)$. CRYSOL 3.0 is designed to produce more accurate results into the WAXS regime than CRYSOL version 2 \cite{franke2017atsas}. Secondly, we independently calculated the scattering in-vacuo, (i.e. $\langle \mid A_a(q)\mid^2 \rangle_\Omega$) using the Debye equation \cite{scardi2016celebrating,debye1915scattering} combined with atomic scattering factors accurate to 6 \AA$^{-1}$ using the method outlined in ref. \cite{waasmaier1995new}. The Debye equation is defined as:
\begin{equation}
    \langle \mid A_a(q)\mid^2 \rangle_\Omega \equiv I_{e u}(q) = N f^2 + f^2\sum_i^N\sum_{i\neq j}^N\frac{\sin(qr_{ij})}{qr_{ij}}
\end{equation}
Where $I_{eu}(q)$ is the calculated scattering intensity in electron units, $r_{ij}$ is the Euclidean distance between the $i^{th}$ and $j^{th}$ atom and $f$ is the atomic scattering factor. The differential scattering cross section ($\frac{d\sigma}{d\Omega}(q)$), which could be directly compared to the experimentally acquired intensity after normalization for the number of molecules in the scattering volume, was obtained through: $\frac{d\sigma_c}{d\Omega}(q)=I_{eu}(q)r_e^2$, where the c subscript indicates in-vacuo calculation and $r_e$ is the electron radius. It should be noted that a slight difference between $\frac{d\sigma_c}{d\Omega}(q)$ and $\frac{d\sigma}{d\Omega}(q)$ exists due to the differences described by equation 1.

 Real-space oligomer structures were constructed using POLYS software \cite{engelsen1996molecular,engelsen2014polys}. The dihedral angles, which define the free-energy minimizing structure were obtained from ref \cite{braccini1999conformational}. The results reported in ref. \cite{braccini1999conformational} agree well with those obtained from fibre diffraction \cite{palmer1945x, wuhrmann1945optik, walkinshaw1981conformations, atkins1973structural,atkins1973structural1, mackie1971conformations}. Where available, structures from the Protein Data Bank were also compared and results are available in the Supplemental Material \cite{sup_mat}.

\begin{figure}[h!]
\includegraphics[width=0.8\columnwidth]{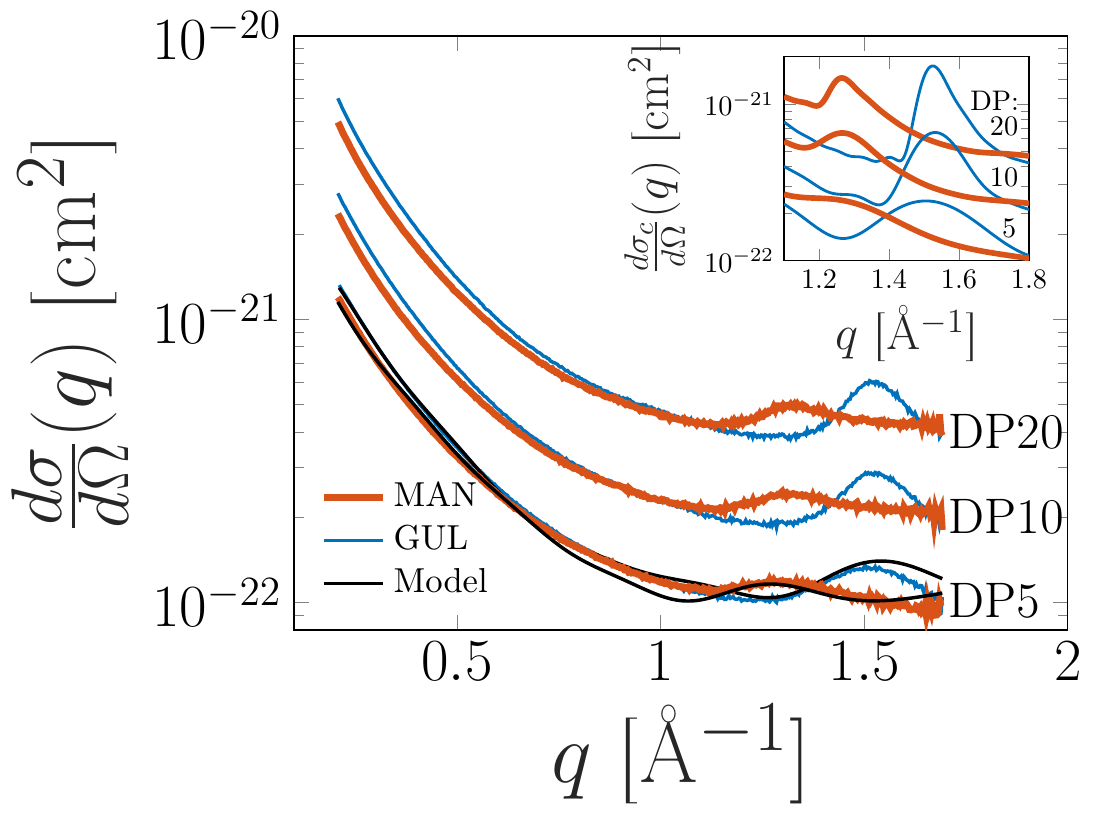}
  \caption{Experimentally measured scattering from $\beta$-D-Man\textit{p}A (MAN) and $\alpha$-L-Gul\textit{p}A (GUL) with differing dps together with the predicted solution scattering for rigid chains calculated using CRYSOL 3.0. The insert shows the predicted in-vacuo scattering calculated using equation 2. The predicted scattering and experimentally measured results agree at dp = 5 and the solution state conformation can be obtained from experimental data. Slight differences can be accounted for from inaccuracies in the calculated hydration layer and a small number of chains populating conformations outside the deepest well in the free energy landscape. At higher dp agreement between the calculated and measured structures requires taking into account flexibility in the glycosidic linkage.}
  \label{fgr:alginate}
\end{figure}

The structures of the three key polyuronates investigated in this letter, at the calculated global free energy minima and in the absence of thermal fluctuations, are shown in figure \ref{fgr:conformers}. It can be seen that the chains conform to a linear arrangement with monomers spaced at different regular intervals along the chain. Such a conformation gives rise to discrete regions of high and low electron density and subsequently a Bragg reflection is predicted experimentally to result from x-ray scattering if the in-vitro solution state conformation of the chains resembles the calculated free-energy-minimizing conformation. Alternatively, significant deviations from this conformation, as a result of thermal fluctuations randomizing the monomer positions, would result in no observed Bragg reflection. The existence of an experimentally-detected Bragg peak at the calculated scattering vector would not only be evidence of the validity of the atomic coordinates used to model the solution state conformation, but would also be reflective of the intrinsic rigidity of the glycosidic backbone. Furthermore, measured changes, or the lack of changes in the peak shape with increasing dp are predicted to reveal the length-scale below which the chains can be considered rigid.
\begin{figure}[h!]
  \includegraphics[width=0.8\columnwidth]{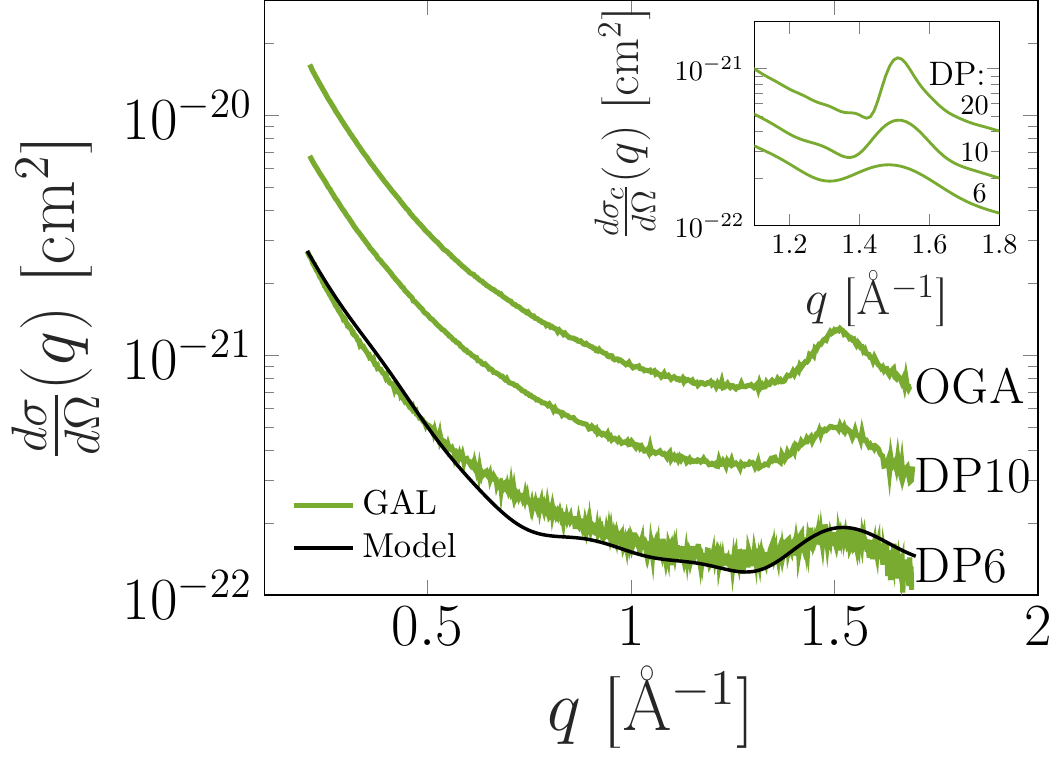}
  \caption{Experimentally measured scattering from $\alpha$-D-Gal\textit{p}A (GAL) with differing dps together with the predicted solution scattering for a rigid chain calculated using CRYSOL 3.0. The insert shows the predicted in-vacuo scattering calculated using equation 2. It can be observed that the calculated solution state scattering agrees with the experimentally obtained signal at dp = 6. Slight differences can be accounted for from inaccuracies in the calculated hydration layer and a small number of chains populating conformations outside the deepest well in the free energy landscape. At higher dp agreement between the calculated conformaiton and experiment requires the inclusion of flexibility in the glycosidic linkage.}
  \label{fgr:galac}
\end{figure}

Figure \ref{fgr:alginate} shows the solution x-ray scattering results for the two alginic acids and figure \ref{fgr:galac} the polygalacturonic acid. It can be seen that a single broad Bragg peak exists in the high-q regime for each species, consistent with the position calculated from the conformers shown in figure \ref{fgr:conformers} using CRYSOL 3.0. (At $0.5\geq q \geq 1$ \AA$^{-1}$ the calculated and experimentally obtained data differs, which we suggest is the result of currently neglecting thermal fluctuations and the small population of other states in the ensemble in the calculation, and this is being investigated in ongoing work employing molecular dynamics.) Figure \ref{fgr:alginate} shows the significant difference in peak position and shape between the results obtained for $\alpha$-L-Gul\textit{p}A and $\beta$-D-Man\textit{p}A. Such a large effect is primarily the result of differences in the geometry of the glycosidic linkage, which is either di-axial for $\alpha$-L-Gul\textit{p}A or di-equatorial for $\beta$-D-Man\textit{p}A. However calculations using chains with glycosidic linkage orientations in the second most populated conformation: a 3(1) helix instead of the most populated 2(1) for $\alpha$-L-Gul\textit{p}A, and a 2(1) helix instead of the most populated 3(1) for $\beta$-D-Man\textit{p}A, showed that the peak positions deviated by 0.1-0.2 \AA$^{-1}$ purely based on the local helical twist irrespective of the linkage type (see the Supplemental Material \cite{sup_mat}). As our data shows, this deviation was clearly visible in the measured peak positions demonstrating that the average free-energy-minimising solution state conformation can be readily obtained with sub {\AA}ngstrom accuracy. 

Figures \ref{fgr:alginate} and \ref{fgr:galac} further show measured changes in the peak shape with increasing dp. It can be seen that at dp=5 and dp=6, respectively, the calculated peak height and width for all three conformers agrees with that obtained experimentally. The $\beta$-D-Man\textit{p}A shows little difference in peak height or width with increasing dp, compared to $\alpha$-L-Gul\textit{p}A and $\alpha$-D-Gal\textit{p}A. This is a result of the increased flexibility of the di-equatorial linked $\beta$-D-Man\textit{p}A, compared to its two more rigid brethren. $\alpha$-L-Gul\textit{p}A and $\alpha$-D-Gal\textit{p}A show increasing peak height and a slight decrease in width at dp=10, above which the peak width and height displays little change compared to the predicted scattering for longer rigid sections of chains (see the inserts of figures \ref{fgr:alginate} and \ref{fgr:galac}). Differences here can be attributed to the flexibility of the glycosidic linkage and are observed above the length-scale at which the average conformation deviates from a rigid linear chain (between dp=6 and dp=10).  As would be predicted the $\alpha$-L-Gul\textit{p}A and $\alpha$-D-Gal\textit{p}A show near identical scattering due to these two conformers being mirror images of each other. 



 The ability to obtain direct structural information regarding the local helical conformation, to sub Angstrom accuracy, in the solution state, is crucial for improving the structural characterization of polysaccharides under conditions not attainable in the condensed state and / or when crystallization is not possible. We have shown that with careful measurements solution state x-ray scattering with synchrotron radiation can resolve atomistic features that reveal the average local helical conformation of the chains, together with information about the intrinsic flexibility of the different linkages.

 The authors thank Benjamin Westberry for assistance with sample transportation and Nigel Kirby for assistance during data acquisition. The NZ synchrotron group are acknowledged for travel funding. Part of this research was undertaken on the SAXS/WAXS beam-line at the Australian Synchrotron, part of ANSTO. Post doctoral funding for BM was provided by the Ministry of Science and Technology, Taiwan under grant No. MOST 105-2221-E-007-137-MY3.





\section*{Supplementary material}

# Supplemental Material for "Polysaccharide conformations measured by solution state x-ray scattering"

Bradley W. Mansel and Hsin-Lung Chen

*Department of Chemical Engineering,*

*National Tsing Hua University, Hsinchu 30013, Taiwan*

Timothy M. Ryan

*Australian Synchrotron, Clayton, VIC 3168, Australia*

Leif Lundin

*RISE, Agrifood and Bioscience, Gothenburg, Sweden*

Martin A.K. Williams

*School of Fundamental Sciences, Massey University,*

*Palmerston North 4442, New Zealand and*

*The MacDiarmid Institute for Advanced Materials*

*and Nanotechnology, Wellington 6140, New Zealand*

**SAMPLE PREPARATION**

SAXS measurements for $\alpha$-D-Gal$p$A consisted of first running sample solutions through an ion-exchange column (Amberlite) to remove any residual ions, and render the galacturonans in their acid form. These were then freeze dried and redissolved at the required concentrations. The pH of solutions was modified to $p$H = 7 using a calibrated pH meter and 0.1 M NaOH to give pH-controlled samples with the lowest salt concentrations possible. All solutions were filtered through 0.2 micron syringe filters before x-ray scattering measurements to remove any contaminants. $\beta$-D-Man$p$A and $\alpha$-L-Gul$p$A were used as received, as information from the supplier showed samples were already in low ionic strength conditions and required no further purification. The near identical scattering of $\alpha$-L-Gul$p$A and $\alpha$-D-Gal$p$A confirmed that both samples had near identical ionic environments.

*Homogalacturonan Oligomers* with degrees of polymerisation (DP) of 6 and 10 were homemade as described in the literature [S1]. Briefly, polygalacturonic acid was subject to a partial digestion with an endo-acting enzyme, endo-PG II, and the digest products were separated from each other using high performance anion exchange chromatography (HPAEC). Fractions containing the polyelectrolyte of interest were pooled from several separations to obtain reasonable quantities.

**POLYGALACTURONIC ACID MODELS EXTRACTED FROM PECTIN METHYLESTERASE CRYSTAL STRUCTURES**

The protein data bank has a number of pectin methylesterase (PME) structures containing homogalacturonan (HG) oligomers. The oligomer structure can be extracted and the coordinates used to calculate the scattering. While the exact conformation might be slightly perturbed by surrounding protein, the structure includes accurate thermal fluctuations. Here we used 2NSP [S2] from the protein data bank. Figure S1 shows the experimental and calculated scattering together with the real-space structure. It can be seen that the peak positions agree with the slight randomization of the monomer positions having little effect on the calculated scattering.

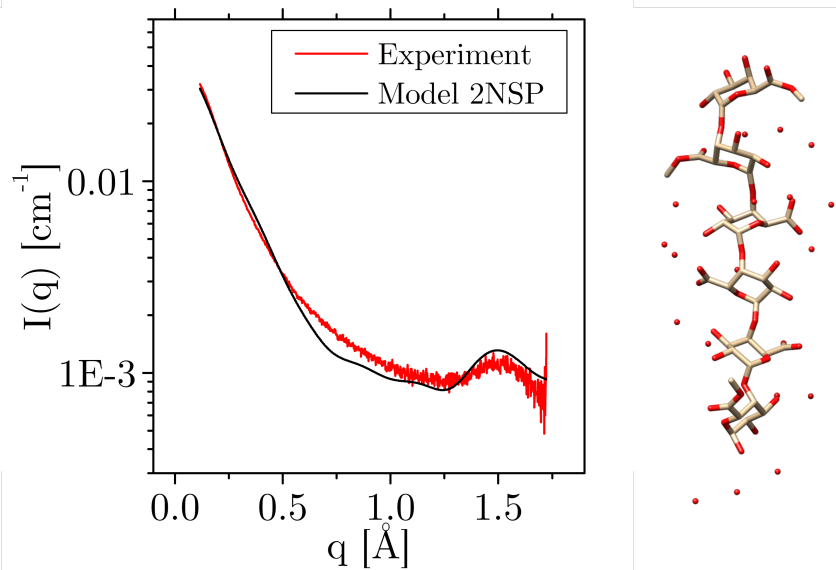

Figure S1. The calculated scattering from the HG contained in the 2NSP crystal structure. The structure shown on the right shows a predominantly 2 fold helix. The calculated scattering, on the left, shows that the calculated and measured peak positions agree. The red spheres in the real-space structure are water molecules which were not included in the calculation.

## SAXS MODELLING FROM ALGINATE CONFORMERS

To confirm that the correct helical conformation of the chains could be obtained from the recorded x-ray scattering signals, modeling was performed with the chains in their second most populated conformation. Figures S2 and S3 show the experimental data together with the fitted chains in different conformations. For the $\alpha$-L-Gul$p$A it can be seen that the peak position of the 2 fold helix agrees with that of the experimental data while the 3 fold helix shows a significant shift to lower q, indicating an expansion of the chain. Indeed, the real-space figure shows that the 3-fold helix is in a more extended conformation agreeing with the calculated profile and the experimental result. $\beta$-D-Man$p$A shows a similar outcome to $\alpha$-L-Gul$p$A where the predicted scattering, namely a 3 fold helix, peak position agrees more favorably with the experiment than the second most likely conformation, the 2 fold helix. Again the second most likely conformation displays a more stretched conformation and the peak also shifts to lower q as expected. The use of wide-angle solution scattering can be

used to obtain the free energy minimizing structure to high accuracy when combined with the atomistic models.

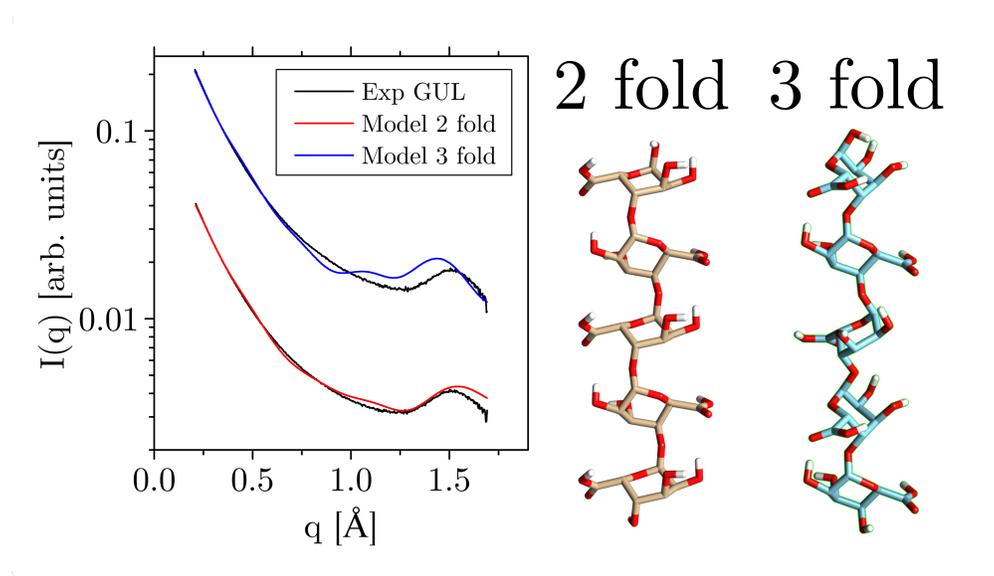

Figure S2. The experimental and calculated scattering from $\alpha$-L-Gul$p$A. The red line shows the model in the most populated conformation, the 2 fold helix, while the blue line represents the second most populated conformation, the 3 fold helix. It can be seen that the peak position of the 2 fold helix shows a better agreement with the experimentally obtained data than the 3 fold helix. The right hand side of the figure shows the two different real-space models used in the calculations, which used CRYSOL 3.0.

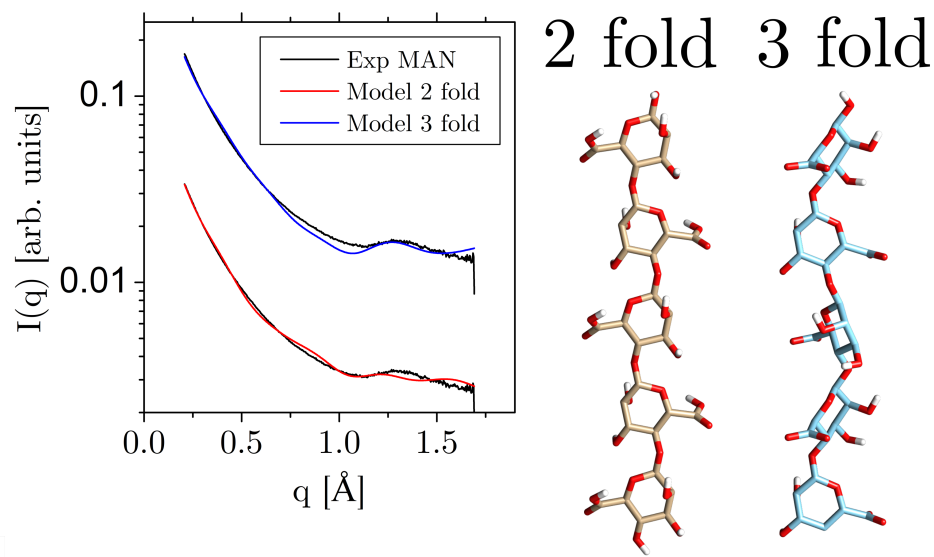

Figure S3. The experimental and calculated scattering from $\beta$-D-Man$p$A. The blue line shows the model in the most populated conformation, the 3 fold helix, while the red line the second most populated conformation a 2 fold helix. It can be seen that the peak position of the 3 fold helix shows a better agreement with the experimentally obtained data than the 2 fold helix. The right hand side of the figure shows the two different real-space models used in the calculations, which used CRYSOL 3.0.

**DETAILS ON DIHEDRAL ANGLES**

Dihedral angles were obtained from ref. [S3] and are listed in table S1. The dihedrals are defined using IUPACIUB commission of Biochemical Nomenclature [S4]

|   | $\alpha$-D-Gal$p$A | $\alpha$-L-Gul$p$A | | $\beta$-D-Man$p$A | |
|---|---|---|---|---|---|
| $\phi$ [°] | **94.5** | **-84.2** | -64.1 | -64.2 | **-85.5** |
| $\psi$ [°] | **147.9** | **-155.3** | -104.4 | -104.6 | **-156.4** |
| $n$ | **2** | **2** | -3 | 2 | **-3** |
| $h$ [Å] | **4.25** | **4.27** | 4.42 | 5.14 | **5.06** |

Table S1. The dihedral angles together with $n$, the number of residues per turn and $h$ the projection of one residue on the helix axis, obtained from ref. [S3]. The bold font indicates the most likely conformation. It can be seen that the most likely conformation corresponds to the smallest $h$ value and reflects our finding in figures S2 and S3. It can also be seen that $\alpha$-D-Gal and $\alpha$-L-Gul$p$A have a similar $h$ value which is also reflected in our experimental findings.



\end{document}